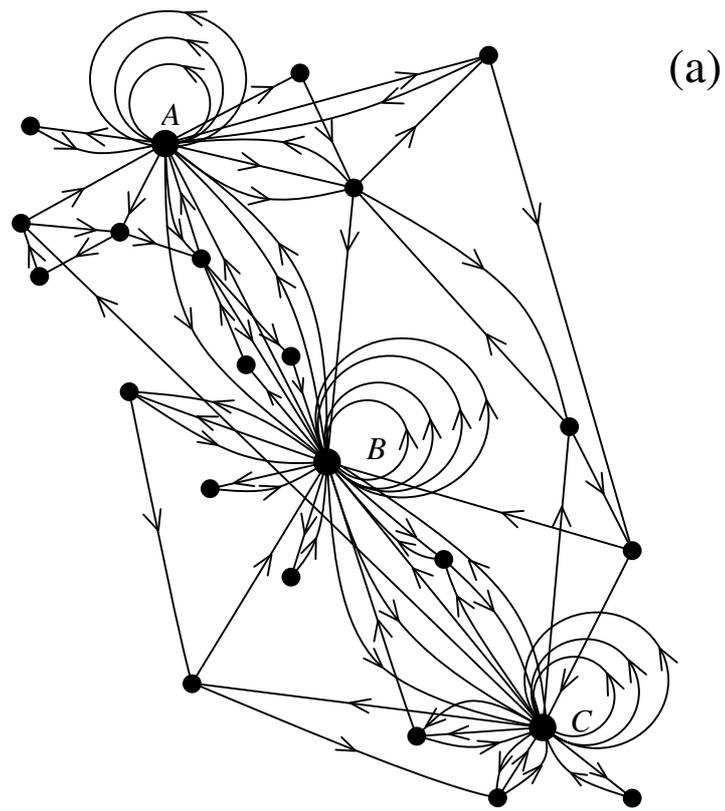

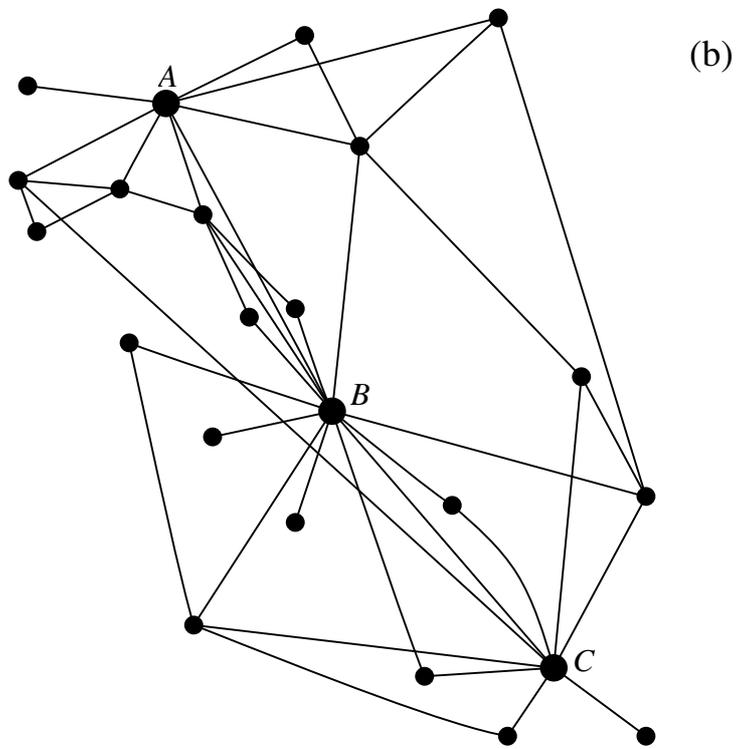

FIG.1

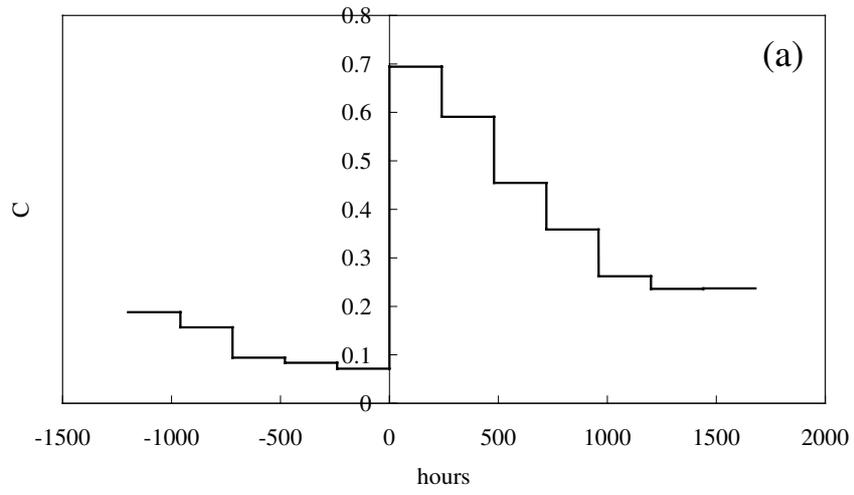
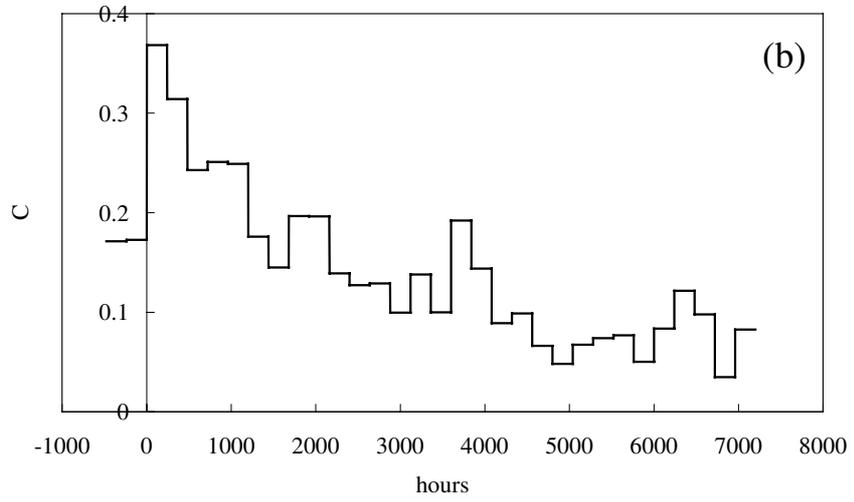
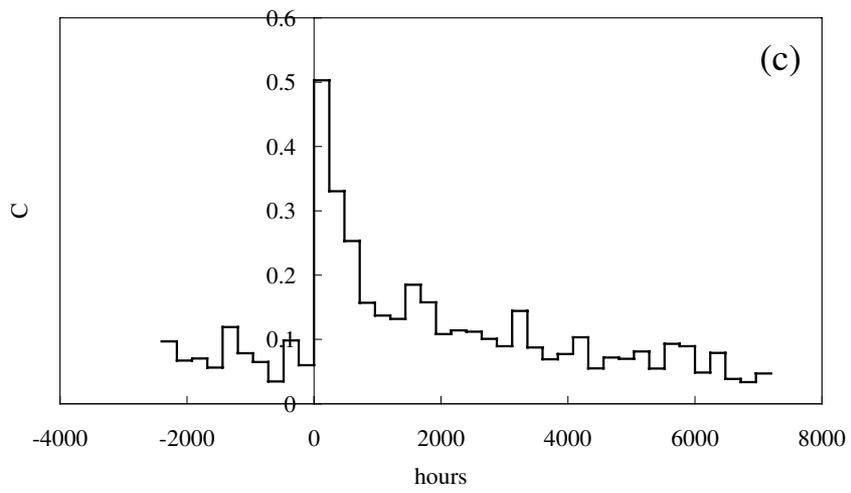

FIG. 2

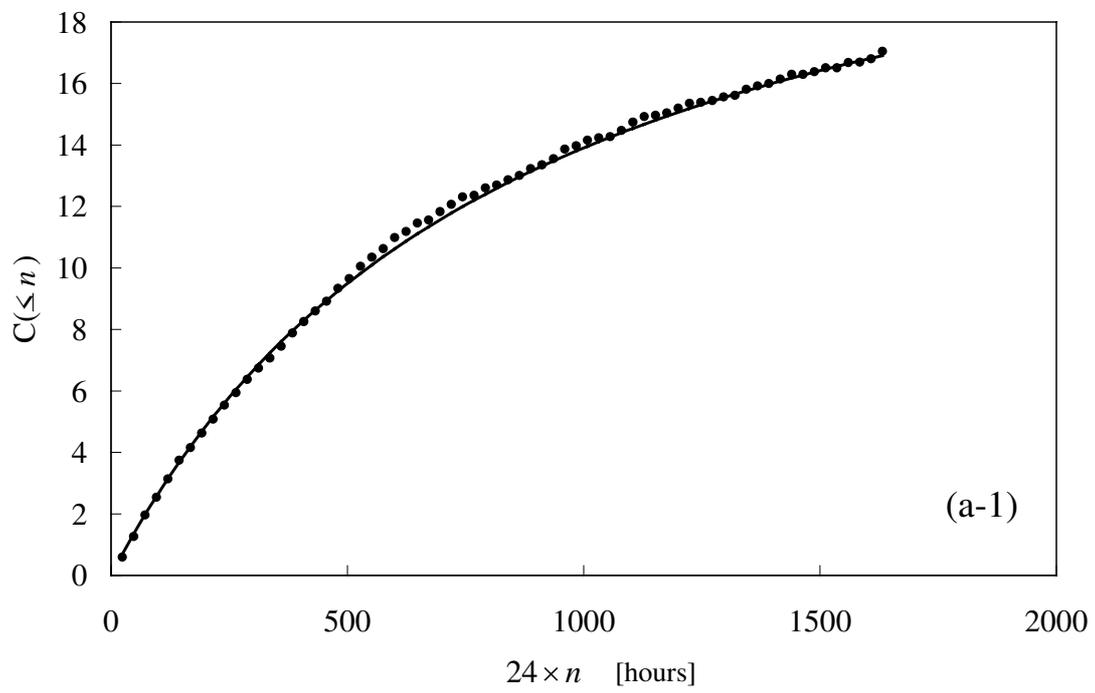

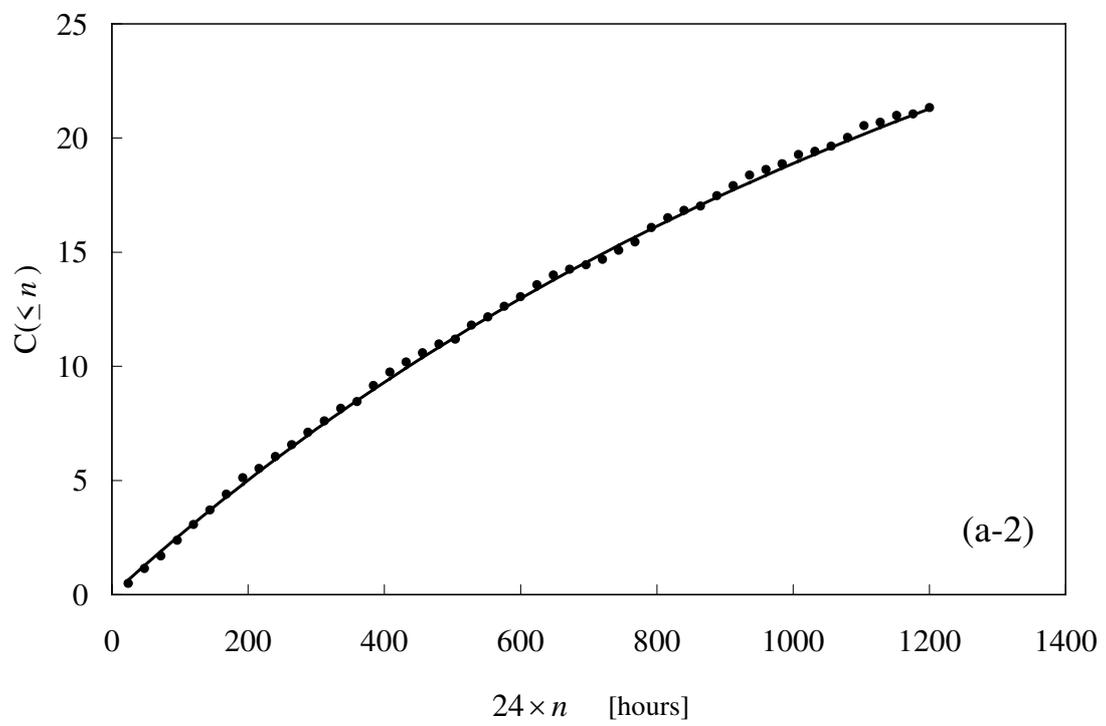

FIG.3(a)

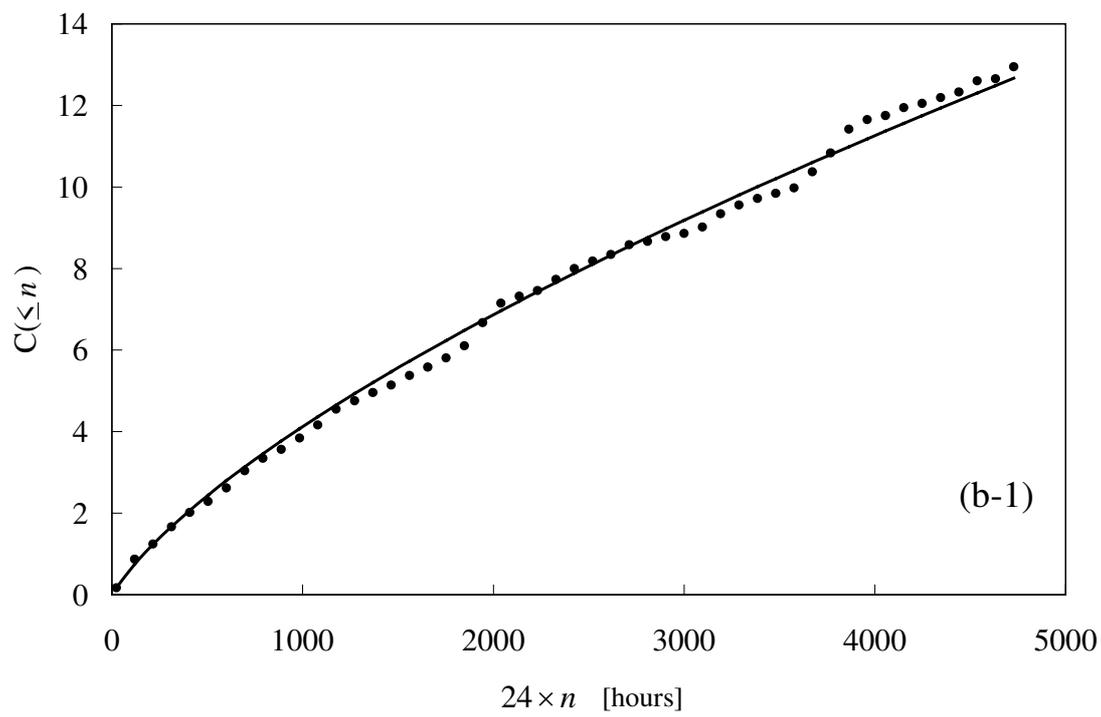

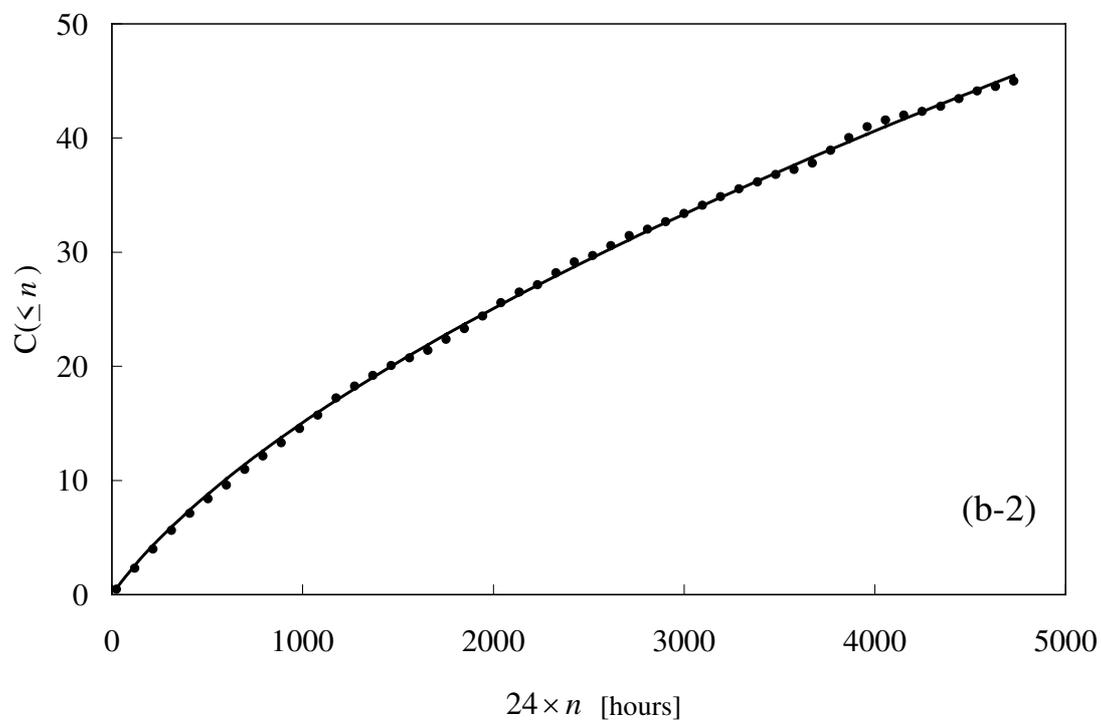

FIG.3(b)

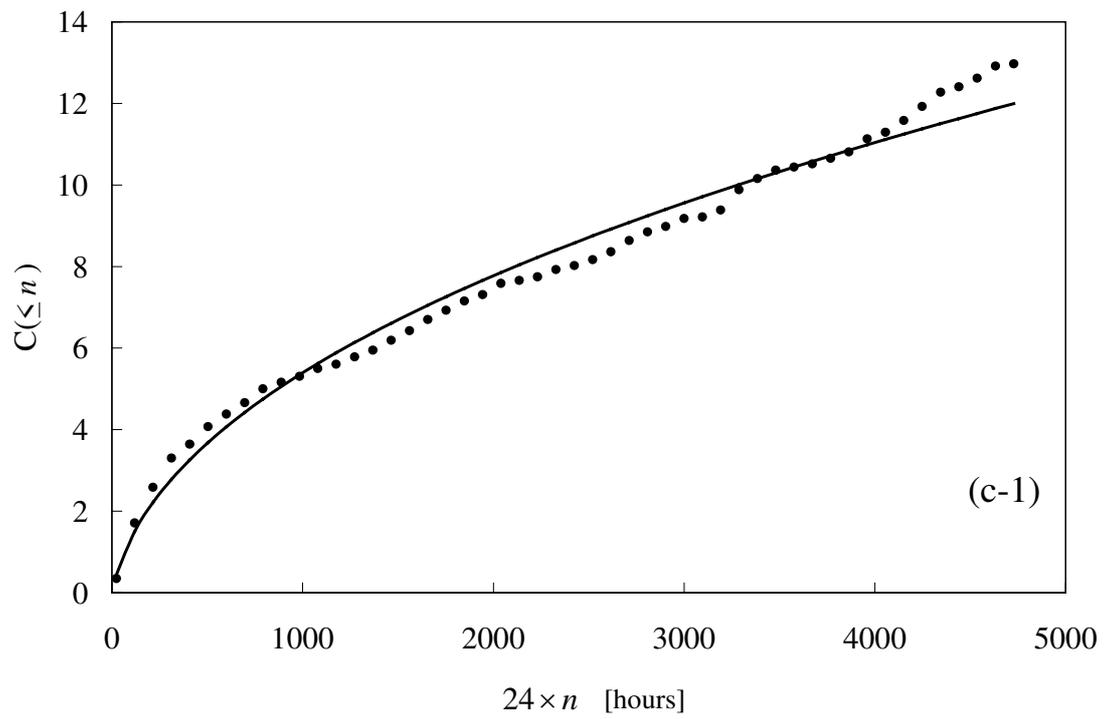

(c-1)

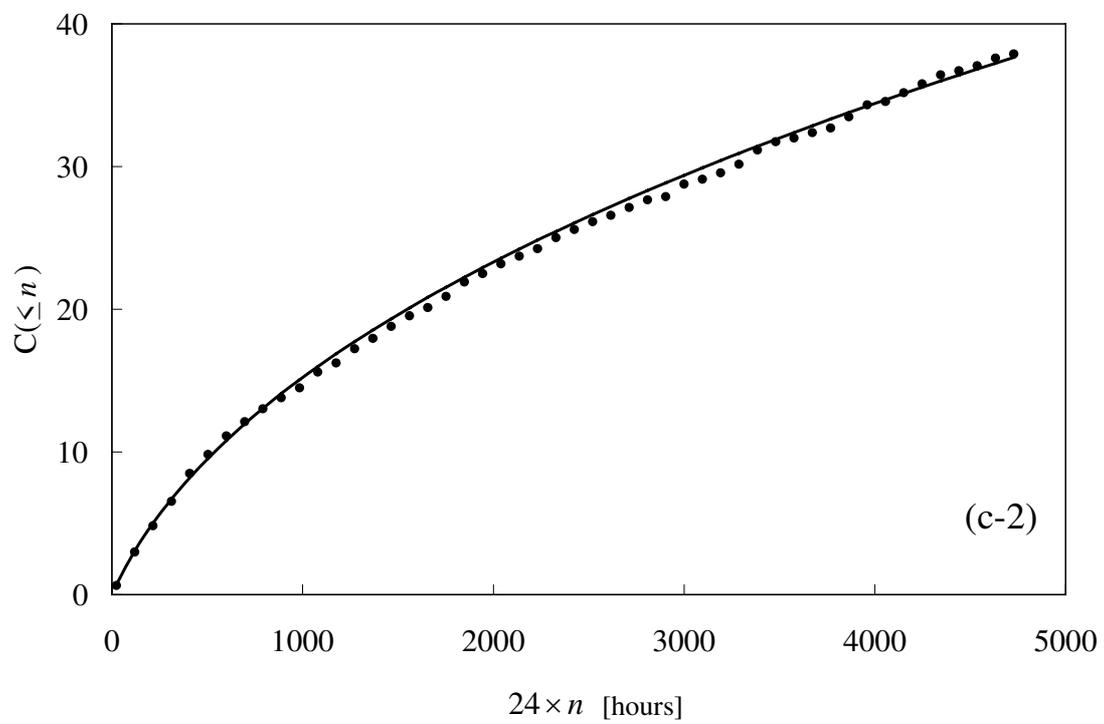

(c-2)

FIG.3(c)

| major event | cell size $(km \times km \times km)$ | $M_0 (\times 10^2)$ | $\alpha$ |
|---|---|---|---|
| Joshua Tree Earthquake | 5 | 9.84 | 2.2 |
| | 10 | 48.0 | 4.0 |
| Landers Earthquake | 5 | 0.618 | 0.33 |
| | 10 | 1.83 | 0.40 |
| Hector Mine Earthquake | 5 | 0.122 | 0.55 |
| | 10 | 1.66 | 0.59 |

TABLE I

# Dynamical evolution of clustering in complex network of earthquakes

Sumiyoshi Abe[1,2] and Norikazu Suzuki[3]

[1]*Institute of Physics, University of Tsukuba, Ibaraki 305-8571, Japan*
[2]*Institut Supérieur des Matériaux et Mécaniques Avancés*,
*44 F. A. Bartholdi, 72000 Le Mans, France*
[3]*College of Science and Technology, Nihon University, Chiba 274-8501, Japan*

**Abstract**   The network approach plays a distinguished role in contemporary science of complex systems/phenomena. Such an approach has been introduced into seismology in a recent work [S. Abe and N. Suzuki, Europhys. Lett. **65**, 581 (2004)]. Here, we discuss the dynamical property of the earthquake network constructed in California and report the discovery that the values of the clustering coefficient remain stationary before main shocks, suddenly jump up at the main shocks, and then slowly decay following a power law to become stationary again. Thus, the network approach is found to characterize main shocks in a peculiar manner.

PACS number(s):   89.75.Da, 91.30.–f, 05.65.+b

Looking at seismic data from the physics viewpoint, it may be of interest to recognize that it is essentially a field-theoretical system. It consists of the series of a set of values of occurrence time, hypocenter, and magnitude of each earthquake. In other words, seismic moment (its logarithm being magnitude) as a field strength is defined on each discrete spacetime point. However, unlike ordinary field dynamics in physics, both the field strength and spacetime points are inherently random. In spite of such apparent complicatedness, known empirical laws are rather simple. There are in fact two celebrated classical examples. One is the Gutenberg-Richter law [1] for the relationship between frequency and seismic moment. The other is the Omori law [2] for the temporal decay of frequency of aftershocks. Both of them are power laws, indicating complexity/criticality of seismicity.

Instantaneous release of huge energy by a main shock can be thought of as a "quenching" process. The disorder of a complex landscape of the stress distribution at faults in the relevant area is then reorganized by it. Accordingly, a swarm of aftershocks may follow. This process constitutes nonstationary parts of a seismic time series, and, due to the power-law nature of the Omori law, "relaxation" to a stationary state is very slow. In a recent work [3], it has been found that there are striking similarities between the aftershock phenomenon and glassy dynamics, including aging and scaling.

In the previous works [4,5], we have studied the spatio-temporal complexity of seismicity and found that both the spatial distance and time interval between two successive earthquakes obey specific but remarkably simple statistical laws. Those results indicate that successive events are indivisibly correlated, no matter how large their spatial separation is. In fact, there is an investigation [6], which points out that an earthquake can be triggered by a foregoing one, which is more than 1000 km away. This implies that the seismic correlation length may be enormously large, exhibiting a strong similarity to phase transition and critical phenomena. Accordingly, it is inappropriate to put spatial windows in analysis of seismicity, in general.

To characterize complexity of event-event correlation in seismicity, we have recently proposed the network approach [7-10], in which seismic data is mapped to a growing random graph. This graph, termed the earthquake network, is constructed as follows. A geographical region under consideration is divided into a lot of small cubic cells. A cell is regarded as a vertex of a network if earthquakes with any values of magnitude occurred therein. Two successive events define an edge between two vertices. If they occur in the same cell, a loop is attached to that vertex. The edges efficiently represent event-event correlation mentioned above. The network thus constructed represents dynamical information of seismicity in a peculiar manner. (Another procedure of constructing an earthquake network, which is more complicated than the present one introducing seven parameters including the spatial distance, time interval, magnitude, and so on, is considered for example in Ref. [11].) Several comments on this construction are in order. Firstly, it contains a single parameter, the cell size, which determines a scale of coarse graining. Once the cell size is fixed, the earthquake network is unambiguously defined. Since there are no *a priori* operational rules to determine the cell size, it is of importance to examine the dependence of the property of earthquake network on it. Secondly, the earthquake network is a directed graph in its nature. Directedness does not bring any difficulties to statistical analysis of connectivity (degree, the number of edges attached to the vertex under consideration) since, by construction, in-degree and out-degree [12] are identical for each vertex with possible exceptions for the first and the last ones in the analysis: that is, the in-degree and out-degree do not have to be distinguished each other in the analysis of connectivity. However, directedness becomes essential when the path length (i.e., the number of edges between a pair of connected vertices) and the period (meaning after how many subsequent earthquakes the event returns to the initial vertex) are considered. Finally, directedness has to be ignored and the path length should be defined as the smallest value among the possible numbers of edges connecting the pair of vertices, when the small-world nature of the earthquake network is investigated. There, loops have to be removed and multiple edges be replaced by single edges. That is, a full directed earthquake network is reduced to a corresponding simple undirected graph (see Fig. 1 for the schematic description).

The earthquake network and its reduced simple graph constructed in this way are found to be scale-free [7] and of the small world [8], exhibit hierarchical organization and assortative mixing [9], and possess the power-law period distributions [10]. A main reason why the earthquake network is heterogeneous is due to the empirical fact that *aftershocks associated with a main shock tend to return to the locus of the main shock, geographically, and therefore the vertices of main shocks play roles of hubs of the network*.

The network approach has been used to examine self-organized-criticality models in the literature [13] if they can reproduce these notable features.

Here, we report a successful application of the dynamical network approach to seismicity. We find through careful analysis that the clustering coefficient exhibits a salient dynamical behavior: it is stationary before a main shock, jumps up at the main shock, and then slowly decays as a power law to become stationary again. We ascertain this behavior for some main shocks occurred in 1990's in California. Thus, the dynamical network approach characterizes a main shock in a peculiar manner.

There are several known quantities that can structurally characterize a complex network. Among them, in particular, we here consider the clustering coefficient introduced in Ref. [14]. This quantity is defined for a simple graph, in which there are no loops and multiple edges contained. A simple graph is conveniently described by the adjacency matrix [15], $A = (a_{ij})$ ($i, j = 1, 2, \cdots, N$ with $N$ being the number of vertices contained in the graph). $a_{ii} = 0$, and $a_{ij} = 1$ (0) if the $i$th and $j$th vertices are connected (unconnected) by an edge. The clustering coefficient, $C$, is then given by

$$C = \frac{1}{N} \sum_{i=1}^{N} c_i, \qquad (1)$$

where

$$c_i = \frac{2 e_i}{k_i (k_i - 1)} \qquad (2)$$

with

$$e_i = (A^3)_{ii} \qquad (3)$$

and $k_i$ the value of connectivity (i.e., the degree) of the $i$th vertex. This quantity has the following meaning. Suppose that the $i$th vertex has $k_i$ neighboring vertices. At most, $k_i(k_i-1)/2$ edges can exist between them. $c_i$ is the ratio of the actual number of edges of the $i$th vertex and its neighbors to this maximum value. Thus, it quantifies the degree of adjacency between two vertices neighboring the $i$th vertex. $C$ is its average over the whole graph. In the earthquake network, $c_i$ quantifies how strongly two aftershocks associated with a main shock (as the $i$th vertex) are correlated.

Now, we address the question as to how the clustering coefficient changes in time as the earthquake network dynamically evolves. For this purpose, we have studied the catalog of earthquakes in California, which is available at URL http://www.data.scec.org/. In particular, we have focused our attention to three major shocks occurred in 1990's: (a) the Joshua Tree Earthquake (M6.1) at 04:50:23.20 on April 23, 1992, 33°57.60'N latitude, 116°19.02'W longitude, 12.33 km in depth, (b) the Landers Earthquake (M7.3) at 11:57:34.13 on June 28, 1992, 34°12.00'N latitude, 116°26.22'W longitude, 0.97 km in depth, and (c) the Hector Mine Earthquake (M7.1) at 09:46:44.13 on October 16, 1999, 34°35.64'N latitude, 116°16.26'W longitude, 0.02 km in depth. We have taken the intervals of the seismic time series containing these events, divided the intervals into many segments, and constructed the earthquake network of each segment. Then, we have calculated the value of the clustering coefficient of each network. In this way, dynamical evolution of clustering has been explored.

In Fig. 2, we present the results on evolution of the clustering coefficient in the case when the length of the segments is fixed to be 240 hours long. Here, the cell size $5\,\text{km}\times 5\,\text{km}\times 5\,\text{km}$ is examined. A remarkable behavior can be appreciated: the clustering coefficient stays stationary before the main shocks, suddenly jumps up at the moments of the main shocks, and then gradually decays.

To clarify the property of the slow decay in more detail, we present Fig. 3, in which shorter-time analysis with 24 hours is performed by examining two different cell sizes, $5\,\text{km}\times 5\,\text{km}\times 5\,\text{km}$ and $10\,\text{km}\times 10\,\text{km}\times 10\,\text{km}$. As can clearly be appreciated, the "cumulative" clustering coefficient,

$$C(\leq n) = \sum_{M=1}^{n} C_M, \tag{4}$$

obeys a definite law, where $C_M$ stands for the clustering coefficient of the network constructed in the interval $24\times (M-1) \sim 24\times M$ [hours] after the moment of the main shock at $M=0$, and $n=(\text{hours})/24$. Indeed, it is well represented by the following power law:

$$C_M \sim \frac{1}{(1+M/M_0)^{\alpha}}, \tag{5}$$

where $\alpha$ and $M_0$ are positive constants, and their values are given in Table I.

In conclusion, we have found that the clustering coefficient of the evolving earthquake network remains stationary before a main shock, suddenly jumps up at the main shock, and then slowly decays to become stationary again following the power-law relaxation. In this way, the clustering coefficient is shown to successfully characterize main shocks. We would like to emphasize that the power-law decay after a main shock described in Eq. (5) might remind one of the Omori law, but actually they are not directly related to each other. This is because, in the definition of the clustering coefficient, loops are removed and multiple edges are replaced by single edges, that is, a number of aftershocks are excluded in the analysis.

One of the authors (S. A.) would like to thank Carmen P. C. Prado for discussions.

# Figure and Table Captions

FIG. 1  Schematic descriptions of an earthquake network. (a) A full directed network. The vertices with high values of connectivity, *A*, *B*, and *C*, correspond to main shocks. (b) The simple undirected graph reduced from the full network in (a).

FIG. 2  Evolution of the (dimensionless) clustering coefficient during each 240 hours. The origins are adjusted to the moments of the main shocks, that is, (a) the Joshua Tree Earthquake, (b) the Landers Earthquake, and (c) the Hector Mine Earthquake.

FIG. 3  Evolution of the (dimensionless) cumulative clustering coefficient defined in Eq. (4) during each 24 hours. The solid curves are due to the model in Eq. (4) with the form in Eq. (5). (a, b, c-1) and (a, b, c-2) are the results for the cell sizes, $5\,\text{km} \times 5\,\text{km} \times 5\,\text{km}$ and $10\,\text{km} \times 10\,\text{km} \times 10\,\text{km}$, respectively, for (a) the Joshua Tree Earthquake, (b) the Landers Earthquake, and (c) the Hector Mine Earthquake. The values of the parameters in Eqs. (5) are given in Table I.

TABLE I  The values of the parameters in Eq. (5) used in Fig. 3.